\documentclass[12pt]{article}
\usepackage{epsfig}
\usepackage{here}
\usepackage{a4}
\usepackage{amssymb}

\begin{document}

\def\Journal#1#2#3#4{{#1} {\bf #2}, #3 (#4)}

\def\etal{{\it et\ al.}}
\def\NCA{\em Nuovo Cim.}
\def\NIM{\em Nucl. Instrum. Methods}
\def\NIMA{{\em Nucl. Instrum. Methods} A}
\def\NPB{{\em Nucl. Phys.} B}
\def\PLB{{\em Phys. Lett.}  B}
\def\PRL{\em Phys. Rev. Lett.}
\def\PRC{{\em Phys. Rev.} C}
\def\PRD{{\em Phys. Rev.} D}
\def\ZPC{{\em Z. Phys.} C}
\def\ASP{{\em Astrop. Phys.}}
\def\JETP{{\em JETP Lett.\ }}

\def\numunue{\nu_\mu\rightarrow\nu_e}
\def\numunutau{\nu_\mu\rightarrow\nu_\tau}
\def\nuebar{\bar\nu_e}
\def\nue{\nu_e}
\def\nutau{\nu_\tau}
\def\numubar{\bar\nu_\mu}
\def\numu{\nu_\mu}
\def\ra{\rightarrow}
\def\numubarnuebar{\bar\nu_\mu\rightarrow\bar\nu_e}
\def\nuebarnumubar{\bar\nu_e\rightarrow\bar\nu_\mu}
\def\osc{\rightsquigarrow}

\def\inteni{{\cal I}_{pot}}
\def\fmerit{{\cal F}}

\thispagestyle{empty}
\vspace*{1cm}
\begin{center}
{\Large{\bf  Experiments For CP-Violation:\\
A Giant Liquid Argon Scintillation, Cerenkov
And Charge Imaging Experiment ?\footnote{Invited talk
at the II International Workshop on: NEUTRINO OSCILLATIONS IN VENICE, Venice (Italy), December 2003.}}}\\
\vspace{.5cm}
{\large A. Rubbia}\footnote{Andre.Rubbia@cern.ch}

Institut f\"{u}r Teilchenphysik, ETHZ, CH-8093 Z\"{u}rich,
Switzerland
\end{center}
\vspace{2.cm}


\abstract{In this paper we address a class of ``ultimate'' generation
experiments for the search of CP-violation in neutrino oscillations. 
Neutrino factories require large magnetized detectors.
New generation superbeams or beta-beams need giant detectors.
The liquid Argon TPC technology has great potentials for both applications.
Although the ICARUS program has demonstrated that this
technology is mature, the possibility to built a giant liquid argon TPC is viewed
by many as a technically impossible and unsafe task. We argue that a
giant liquid argon Cerenkov and charge Imaging experiment
would be an ideal match for a superbeam or a betabeam.
Such a detector would in addition cover a broad
physics program, including the observation of atmospheric neutrinos, 
solar neutrinos, supernova neutrinos, and search for
proton decays, in addition to the accelerator physics program.
We show a potential implementation of such a giant LAr detector and argue
that it could be technically feasible. The possibility to host such a detector
in an underground cavern is under study.
}


\newpage
\pagestyle{plain} 
\setcounter{page}{1}
\setcounter{footnote}{0}

\normalsize\baselineskip=15pt

\section{Introduction}
How can one experimentally observe the CP-violation in the leptonic sector? From the 
unitary mixing matrix, which can be parameterized as
\begin{equation}
U(\theta_{12},\theta_{13},\theta_{23},\delta)=\left(
\begin{tabular}{ccc}
$c_{12}c_{13}$      & $s_{12}c_{13}$   &  $s_{13}e^{-i\delta}$ \\
$-s_{12}c_{23}-c_{12}s_{13}s_{23}e^{i\delta}$ &
$c_{12}c_{23}-s_{12}s_{13}s_{23}e^{i\delta}$ & $c_{13}s_{23}$ \\
$s_{12}s_{23}-c_{12}s_{13}c_{23}e^{i\delta}$ &
$-c_{12}s_{23}-s_{12}s_{13}c_{23}e^{i\delta}$ & $c_{13}c_{23}$ 
\end{tabular}\right)
\end{equation}
with $s_{ij}=\sin\theta_{ij}$ and $c_{ij}=\cos\theta_{ij}$,
we get the freedom of the complex phase
(physical only if $\theta_{13}\neq 0$ !). We know that 
\begin{enumerate}
\item the $\delta$-phase can only be observed in an appearance experiment since
the disappearance is a T-symmetric process;
\item the effect for antineutrinos should be opposite to neutrinos ($\delta\rightarrow -\delta$);
\item it should have the expected $L/E_\nu$ dependence, where $E_\nu$ is
the neutrino energy;
\end{enumerate}
Considering oscillations involving electron and muon flavors, the oscillation probability, in 
the parameterization described above, is:
\begin{eqnarray}
 P(\nu_e\ra\nu_\mu)   = P(\bar\nu_\mu\ra\bar\nu_e)  = \nonumber \\ 
4c^2_{13}
\Bigl[  \sin^2\Delta_{23} s^2_{12} s^2_{13}
      s^2_{23} + c^2_{12}
      \left( \sin^2\Delta_{13}s^2_{13}s^2_{23} + 
        \sin^2\Delta_{12}s^2_{12}
         \left( 1 - \left( 1 + s^2_{13} \right) s^2_{23} \right)  \right)  \Bigr]  
\nonumber \\ 
  -  \frac{1}{2}c^2_{13}\sin (2\theta_{12})s_{13}\sin (2\theta_{23}) \cos\delta 
     \left[ \cos 2\Delta_{13} - \cos 2\Delta_{23} - 
       2\cos(2\theta_{12})\sin^2\Delta_{12}\right]  \nonumber \\ 
  + \frac{1}{2}c^2_{13}\sin\delta 
\sin(2\theta_{12})s_{13}\sin (2\theta_{23})
\left[ \sin2\Delta_{12} - \sin2\Delta_{13} + 
\sin 2\Delta_{23} \right]
\end{eqnarray}
where $\Delta_{jk}\equiv
\Delta m^2_{jk}L/4E_\nu$ (in natural units).

We note that a precise measurement of
the $\nue\ra\numu$ oscillation probability can yield information
of the $\delta$-phase provided that the other oscillation
parameters in the expression are known sufficiently accurately.
In practice, one can introduce a certain number of discriminants\cite{Bueno:2001jd,Rubbia:2001pk}
which are good quantities to experimentally search for a non-vanishing phase:
\begin{enumerate}
\item $\Delta_\delta \equiv  P(\nu_e\ra\nu_\mu,\delta=+\pi/2)-
P(\nu_e\ra\nu_\mu,\delta=0)$\\
The discriminant $\Delta_\delta$ can be used in an experiment where
one is comparing the measured $\nue\ra\numu$ oscillation probability as a
function of the neutrino energy $E_\nu$ with a ``Monte-Carlo
prediction'' of the spectrum in absence of $\delta$-phase.
\item $\Delta_{CP}(\delta) \equiv  P(\nu_e\ra\nu_\mu,\delta)-P(\bar\nu_e\ra\bar\nu_\mu,\delta)$ \\
The discriminant $\Delta_{CP}$ can be used in an experiment by
comparing the appearance of negative and positive muons as a function of the
neutrino energy $E_\nu$ for $\nue$ and $\bar\nue$ sources.
\item $\Delta_{T}(\delta) \equiv
P(\nu_e\ra\nu_\mu,\delta)-P(\nu_\mu\ra\nu_e,\delta)$ or
$\bar\Delta_{T}(\delta) \equiv  P(\bar\nu_e\ra\bar\nu_\mu,\delta)-
P(\bar\nu_\mu\ra\bar\nu_e,\delta)$ \\
The discriminant $\Delta_{T}$ can be used in an experiment by
comparing the appearance of $\nu_\mu$  {\bf and}
$\nu_e$ (resp. $\bar\nu_\mu$  {\bf and} $\bar\nu_e$) as a function of the
neutrino energy $E_\nu$.
\end{enumerate}
Each of these discriminants have their advantages and disadvantages\cite{Bueno:2001jd}.
The $\Delta \delta$-method can provide excellent determination of the phase limited
only by the statistics of accumulated events, in practice, systematic
effects will have to be carefully kept under control in order to look for a
small effect in a seen-data versus Monte-Carlo-expected comparison.
In addition, the precise knowledge of the other oscillation parameters will
be important.
The $\Delta_{CP}$ is quite straight-forward, it suffers, however, from matter effects
which ``spoil'' $\Delta_{CP}$ since it involves both neutrinos and
antineutrinos, which oscillate differently
through matter. Hence, the $\Delta_{CP}$ requires a good understanding of
the effects related to matter. These effects increase with baseline and
are maximum around the ``resonance energy'' around $E_\nu\simeq 10\ GeV$\cite{Bueno:2001jd}.
The $\Delta_{T}$ is the theoretically cleanest method, since it
does not suffer from the problems of $\Delta \delta$ and
$\Delta_{CP}$. Indeed, a difference in oscillation probabilities
between $\nue\ra\numu$ and
$\numu\ra\nue$ would be a direct proof
for a non-vanishing $\delta$-phase. In addition, matter affects both
probabilities in a same way, since it involves only neutrinos (resp. antineutrinos for $\bar\Delta_{T}$).

\section{Guidelines for a CP-violation neutrino oscillation experiment}
A long-baseline neutrino oscillation experiment designed to search for a non-vanishing
$\delta$-phase is necessarily an ``ultimate'' (or at least a ``phase-II'') experiment
since it should
\begin{itemize}
\item be designed to have ample statistics (for a given $\theta_{13}$) to precisely
determine the oscillation probability as a function of the neutrino energy; hence its size
will depend on the actual value of $\theta_{13}$ (and if $\theta_{13}>0$ !)
\item have an excellent energy resolution to observe the energy dependence of
the oscillation probability and help lift degeneracy of the parameters governing
the neutrino oscillations (see e.g. Ref.\cite{Bueno:2001jd});
\item be performed at a wide-band neutrino beam to cover enough ``oscillations'' peaks
or do ``counting'' at different neutrino beam energy settings;
\item have the possibility to study neutrinos and antineutrinos ideally separately
in order to lift degeneracies (even in the counting mode).
\end{itemize}
\begin{figure}[htb]
\centering
\epsfig{file=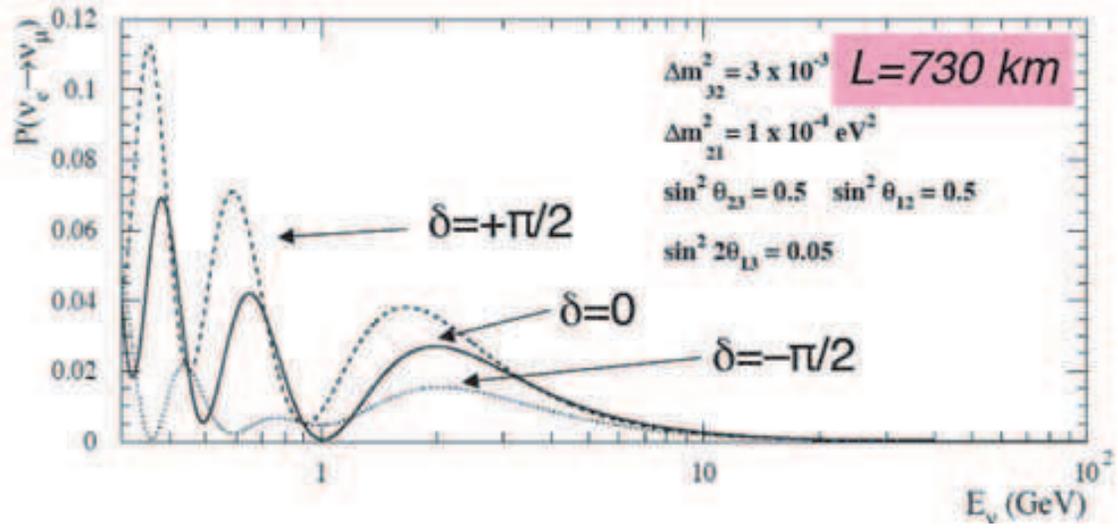,width=15cm}
\caption{The $\numu\rightarrow\nue$ oscillation probability as a function of neutrino
energy for different values of the $\delta$ phase for a given set of oscillation 
parameters and a distance of 730 km.}
\label{fig:oscillgraph}
\end{figure}
Figure~\ref{fig:oscillgraph} illustrates qualitatively the fact that a measurement of the oscillation
probability {\it as a function of energy with good resolution} would indeed
provide direct information on the $\delta$-phase, since this latter introduces
a well-defined energy dependence of the oscillation probability, which is different
from the, say, energy dependence introduced by $\theta_{13}$ alone (when $\delta=0$).

To study these oscillations, one considers three types of neutrino beams produced at accelerators:
\begin{itemize}
\item \underline{Superbeams}: these are conventional pion beams of high intensity, where
the sign of the focusing can be selected to enhance neutrinos or antineutrinos
components
\begin{equation}
\pi^+\rightarrow \mu^++\numu\ \ \ \ \ or\ \ \ \ \ \pi^-\rightarrow \mu^-+\bar\numu
\end{equation}
\item \underline{Betabeams}\cite{Zucchelli}: these are radioactive storage rings, where the type of ion
can be chosen to select electron-neutrinos or electron-antineutrinos
\begin{equation}
~^Z A \rightarrow ~^{Z-1} A \beta^+ \nue\ \ \ \ \ or\ \ \ \ \ ~^Z A \rightarrow ~^{Z+1} A \beta^- \bar\nue
\end{equation}
\item \underline{Neutrino factories}\cite{Geer:1997iz}: these are muon storage rings, where the charge of the muons
can be a priori chosen to select neutrinos/antineutrinos of given flavors-pairs
\begin{equation}
\mu^-\rightarrow e^-\bar\nue\numu\ \ \ \ \ or\ \ \ \ \ \mu^+\rightarrow e^+\nue\bar\numu
\end{equation}
\end{itemize}
From the point of view of neutrino oscillation studies, these beams turn out to be
very complementary. Indeed, the combination of superbeams ($\pi^\pm$) and beta-beams  ($\beta^\pm$) 
allows to study CP- and T-violation by a direct comparison of the results from
the various sources:\\
\centerline{\epsfig{file=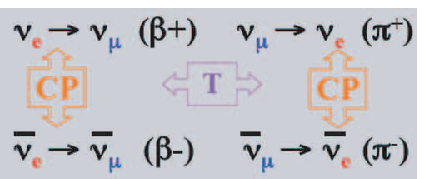,height=3cm}}
In a similar way, a neutrino factory  ($\mu^\pm$)  can provide all possible combinations
to access directly CP- and T-violation:\\
\centerline{ \epsfig{file=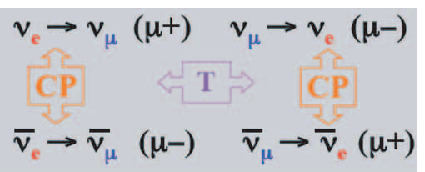,height=3cm}}

However, it goes without saying that these types of beams require very different 
accelerator technologies. Also from a detector point of view, there are quite
distinct optimizations required for super- and beta-beams on one side and
neutrino factory on the other, as discussed in the following sections. In particular,
\textit{detectors for a neutrino factory are necessarily magnetized, hence this will limit
their size to say 10-50~kton, while detectors for super- and beta-beams are conceived
as giant non-magnetized experiments with sizes of say 100-1000~kton.}

\subsection{Detectors for the super and beta-beams}
Superbeams and betabeams are typically designed with energies in the
range of the GeV or below. Since the cross-sections are approximately
linear with energy,  detectors
must be very massive, typically in the range of 100-1000~kton. 
For the superbeams, a good $e/\pi^0$ discrimination
is important in order to suppress the neutral current background with
a leading $\pi^0$:
\begin{equation}
\nu N \rightarrow \pi^0 + X
\end{equation}
from the oscillation channel
\begin{equation}
\numu \rightarrow \nue\ \ \ and\ \ \ \nue N \rightarrow e + X
\end{equation}
In the case of the betabeams, a good $\mu/\pi^\pm$ discrimination
is important in order to suppress the neutral current background with
a charged leading $\pi^\pm$:
\begin{equation}
\nu N \rightarrow \pi^\pm + X
\end{equation}
from the oscillation channel
\begin{equation}
\nue \rightarrow \numu\ \ \ and\ \ \ \numu N \rightarrow \mu + X
\end{equation}
For the beta-beam there is a minimum baseline $L_{min}$ between the source and
the detector. Indeed, the oscillation is detected
through the muon appearance channel with has an energy 
threshold of about 110~MeV. Hence, 
\begin{equation}
L_{min} = \pi\frac{110 MeV}{1.27\Delta m^2}\approx 136\rm\ km
\end{equation}
for $\Delta m^2=2\times 10^{-3}\ eV^2$.

Summarizing, the {\it ideal detector at the superbeam or beta-beam should possess the
following characteristics}:
\begin{itemize}
\item {\bf Low energy threshold}: the detector should be able to detect, reconstruct
and analyse events with neutrino energies in the GeV and below.
\item {\bf Particle identification}: the detector should be able to identify and
measure electrons and muons, and separate them from other hadrons
(typ. neutral and charged pions).
\item {\bf Energy resolution}: the incoming neutrino energy $E_\nu$ is
reconstructed as $E_\nu = E_\ell+E_{had}$, where $E_\ell$ is the leading
lepton energy and $E_{had}$ is the hadronic energy. Hence, detector with
better energy resolution will reconstruct the parameter of the incoming
neutrino better, and therefore the oscillation probability\footnote{Note that
it is often said that Fermi motion spoils energy resolution at low energy. This is not
true when the final state is measured completely. Fermi motion can introduce a high
momentum imbalance (up to 200~MeV), however, the energy available is
related to the binding energy in the nucleus and is hence small compared to the
incoming neutrino energy.}.
\item {\bf Isotropy}: at low energy final states particles are produced at all angle,
in particular, the leading lepton can even be produced backward. This is increased
on a nuclear target by the Fermi motion. The probably most efficient way to detect
these particles is to build a
large neutrino detector, {\it isotropic in nature}, capable of measuring
equally well particles at all angles. This is also a necessary condition
for studying astrophysical sources and look for proton decay.
\end{itemize}
We note that a magnetic field is not necessarily mandatory for these experiments.

\subsection{Detectors for the neutrino factory}
Neutrino factories are typically designed with energies much higher
than the GeV. Given energies in the range of 10~GeV, detectors
are typically in the range of 10-50~kton. 
At high energy, one could independently study the following flavor transitions:
\begin{eqnarray}
\mu^-\ra e^-& \bar\nue&\numu \nonumber \\ 
& & \ra \nue\ra e^-\rm \ appearance\\
& & \ra \numu\rm \ disappearance, \ same \ sign \ muons\\
& & \ra \nutau\ra\tau^-\rm \ appearance, \ high \ energy \ nu's\\
 & \ra &\bar\nue\rm \ disappearance\\
 & \ra &\bar\numu\ra\mu^+\rm \ appearance, \ wrong \ sign \ muons\\
 & \ra &\bar\nutau\ra\tau^+\rm \ appearance, \ high \ energy \ nu's
\end{eqnarray}
plus 6 other charge conjugate processes initiated
from $\mu^+$ decays.
{\it The ideal neutrino detector should be able to measure these 12
different processes as a function of the baseline $L$ and of the neutrino
energy $E_\nu$!}

Of particular interest are the charged current neutrino interactions, since
they can in principle be used to tag the neutrino flavor and helicity,
through the detection and identification of the final state charged lepton:
\begin{eqnarray}
\nu_\ell N \ra \ell^-+hadrons\ \ \ \ \ \bar\nu_\ell N\ra\ell^++hadrons
\end{eqnarray}

Hence, the {\it ideal detector at the neutrino factory should possess the
following characteristics}:
\begin{itemize}
\item {\bf Particle identification}: the detector should be able to identify and
measure the leading charged lepton of the interaction, in order to tag the
incoming neutrino flavor.
\item {\bf Charge identification}: the sign of the leading lepton charge should
be measured, since it tags the helicity of the incoming neutrino. The detector
must necessarily be magnetized.
\item {\bf Energy resolution}: the incoming neutrino energy $E_\nu$ is
reconstructed as $E_\nu = E_\ell+E_{had}$, where $E_\ell$ is the leading
lepton energy and $E_{had}$ is the hadronic energy. Hence, detector with
better energy resolution will reconstruct the parameter of the incoming
neutrino better, and therefore the oscillation probability.
\item {\bf Isotropy}: one might want to perform various similar experiments at different
baselines. The probably most efficient way to achieve this is to build a
large neutrino detector, {\it isotropic in nature}, capable of measuring
equally well
neutrinos from different sources located at different baselines
$L$. Because of the spherical shape of the Earth, sources located at different
baselines $L$ will reach the detector ``from below'' at different angles.
Isotropy of reconstruction is also a necessary condition
for studying astrophysical sources and look for proton decay.
\end{itemize}

\section{The liquid Argon Technology}
The technology of the Liquid Argon Time Projection Chamber (LAr TPC), first
proposed by C.~Rubbia in 1977~\cite{intro1}, was conceived as a tool for
a completely uniform imaging with high accuracy of 
massive volumes.
The operational principle of the LAr TPC is based on
the fact that in highly purified LAr ionization tracks can be transported practically
undistorted by a uniform electric field over macroscopic distances. Imaging is
provided by a suitable set of electrodes (wires) placed at the end of
the drift path continuously sensing and recording the signals induced by
the drifting electrons. 

Non--destructive read--out of ionization electrons by
charge induction allows to detect the signal of electrons crossing
subsequent wire planes with different orientation. This provides several
projective views of the same event, hence allowing space point 
reconstruction and precise calorimetric measurement.
%


The main technological challenges of this detection technique have been recently summarized
elsewhere\cite{t600paper}. They mainly consisted in: (1) techniques of Argon purification (2) operation of wire chambers in cryogenic liquid
and without charge amplification (3) extremely low-noise analog electronics (4) continuous wave-form recording
and digital signal processing.

The feasibility of the technology has been
demonstrated by the extensive ICARUS R\&D programme, which included ten years
of studies on small LAr volumes (proof of principle, LAr purification
methods, read--out schemes, electronics) and five years of studies with
several prototypes of increasing mass (purification technology,
collection of physics events, pattern recognition, long duration tests,
read--out technology). The largest of these devices had a mass of 3 tons of
LAr~\cite{3tons,Cennini:ha} and has been operated continuously for more than four years, collecting
a large sample of cosmic--ray and gamma--source events. Furthermore, a smaller
device (50 l of LAr~\cite{50lt}) was exposed to the CERN neutrino
beam, demonstrating the high recognition capability of the technique for
neutrino interaction events.

\begin{figure}[htb]
\centering
\epsfig{file=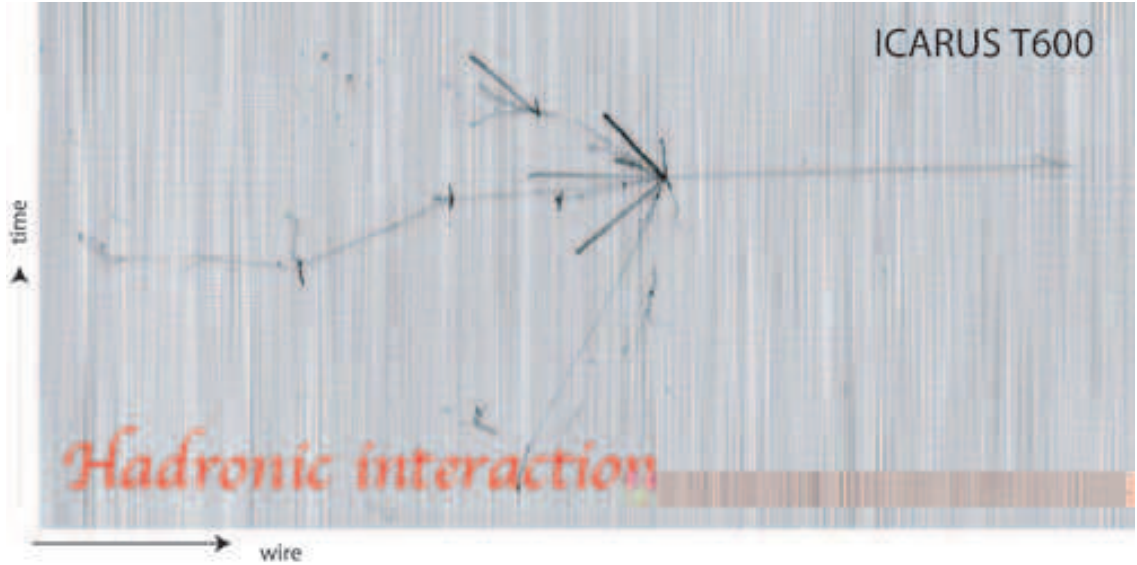,width=15cm}
\caption{Hadronic interaction collected during the ICARUS T600 technical run on the surface.}
\label{fig:hadrointeract}
\end{figure}

The realization of the T600 detector (from design to construction) lasted
about four years and culminated with a full test of the experimental set--up,
carried out at surface during 2001. This test demonstrated the maturity of the project.
All technical aspects of the system, namely cryogenics, LAr purification,
read--out chambers, detection of LAr scintillation light, electronics and DAQ had
been tested and performed as expected. Statistically significant samples of
cosmic--ray events (long muon tracks, hadronic interactions (see e.g Figure~\ref{fig:hadrointeract}),
spectacular high--multiplicity muon
bundles, electromagnetic and hadronic showers, low energy events) were
recorded. The analysis of these events has allowed the development and fine
tuning of the off--line tools for the event reconstruction and the extraction
of physical quantities. It has also demonstrated the performance of
the detector in a quantitative way.

The detector performance can be summarized as:
\begin{itemize}
\item a tracking device with precise event topology reconstruction
\item momentum estimation via multiple scattering
\item measurement of local energy deposition ($dE/dx$), providing
$e/\pi^0$ separation (sampling typ. $2\%X_0$), particle identification
via range versus $dE/dx$ measurement
\item total energy reconstruction of the event from charge integration 
(the volume can be considered as a full-sampling, fully homogenous calorimeter)
providing excellent accuracy for contained events
\end{itemize}
The energy resolutions are:
\begin{itemize}
\item $\sigma/E = 11\%/\sqrt{E(MeV)}\oplus 2\%$ for low energy electrons (measured\cite{Amoruso:2003sw})
\item $\sigma/E \approx 3\%/\sqrt{E(GeV)}$ for electromagnetic showers
\item $\sigma/E \approx 30\%/\sqrt{E(GeV)}$ for hadronic  showers (pure LAr)
\item $\sigma/E \approx 17\%/\sqrt{E(GeV)}$ for hadronic  showers (TMG doped LAr)
\end{itemize}

It is fair to say that the technique has reached a high level of maturity.

\section {A magnetized Liquid Argon TPC}
Liquid argon imaging provides very good
tracking with $dE/dx$ measurement, and excellent calorimetric 
performance for contained showers. This allows for a very
precise determination of the energy of the particles in
an event. This is particularly true for electron
showers, which energy is very precisely measured. 

\begin{figure}[htb]
\centering
\epsfig{file=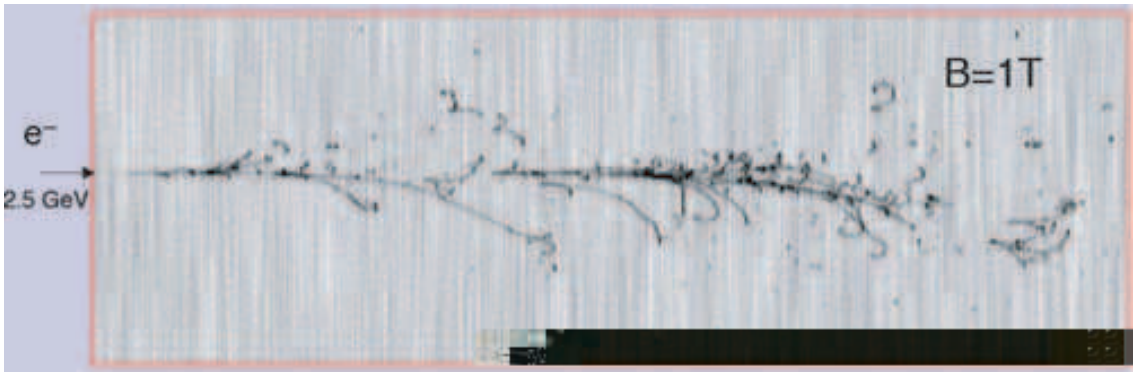,width=15cm}
\caption{Magnetized liquid argon TPC: simulation of the 2.5 GeV electron
shower in liquid argon. The field has a strength B=1~T and is directed
perpendicular to the sheet-plane.}
\label{eshower}
\end{figure}

The possibility to complement these features with those
provided by a magnetic field has been considered\cite{Rubbia:2001pk} and would
open new possibilities:
\begin{itemize}
\item charge discrimination
\item momentum measurement of particles escaping the detector (e.g. high energy muons)
\item very precise kinematics, since the measurements are multiple scattering
dominated (e.g. $\Delta p/p\simeq 4\%$ for a track length of $L=12\ m$ and
a field of $B=1T$).
\end{itemize}
The orientation of the magnetic field is such that the bending direction
is in the direction of the drift where the best spatial resolution is achieved
(e.g. in the ICARUS T600 a point resolution of $400\ \mu m$ was obtained).
The magnetic field is hence perpendicular to the electric field. The Lorentz
angle is expected to be very small in liquid (e.g. $\approx 30 mrad$ at
$E=500\ V/cm$ and $B=0.5T$).
Embedding
the volume of argon into a magnetic field would therefore not
alter the imaging properties of the detector and
the measurement of the bending of
charged hadrons or penetrating
muons would allow a
precise determination of the momentum and 
a determination of their charge.

\begin{figure}[htb]
\centering
\epsfig{file=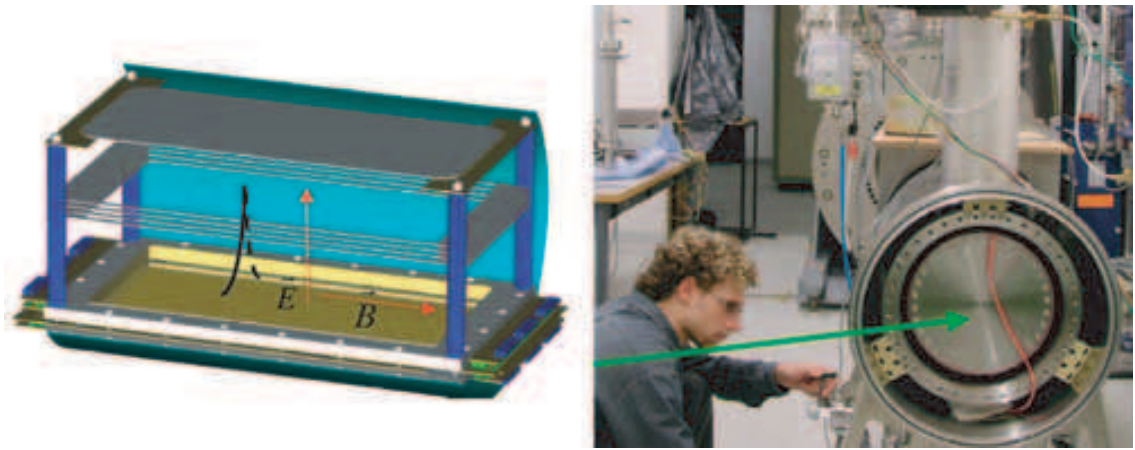,width=15cm}
\caption{Magnetized liquid argon TPC: ongoing R\&D to test liquid argon imaging in a
magnetic field. (left) CAD drawing of the chamber; the direction of the fields
and a bending track is shown (right) actual setup. }
\label{fig:laffranchi}
\end{figure}

The required magnetic field for charge discrimination for a path $x$ in
the liquid Argon is given by the bending
\begin{equation}
b\approx \frac{l^2}{2R}=\frac{0.3B(T)(x(m))^2}{2p(GeV)}
\end{equation}
and the multiple scattering contribution:
\begin{equation}
MS\approx \frac{0.02(x(m))^{3/2}}{p(GeV)}
\end{equation}
The requirement for a $3\sigma$ charge discrimination can be written as:
$b^+-b^- = 2b > 3MS$, which implies:
\begin{equation}
B\geq \frac{0.2(T)}{\sqrt{x(m)}}
\end{equation}
For long penetrating tracks like muons, a field of $0.1T$ allows
to discriminate the charge for tracks longer than 4 meters. This
corresponds for example to a muon momentum threshold of 800~MeV/c.
Hence, performances are very good, even at very low momenta.

\begin{figure}[htb]
\centering
\epsfig{file=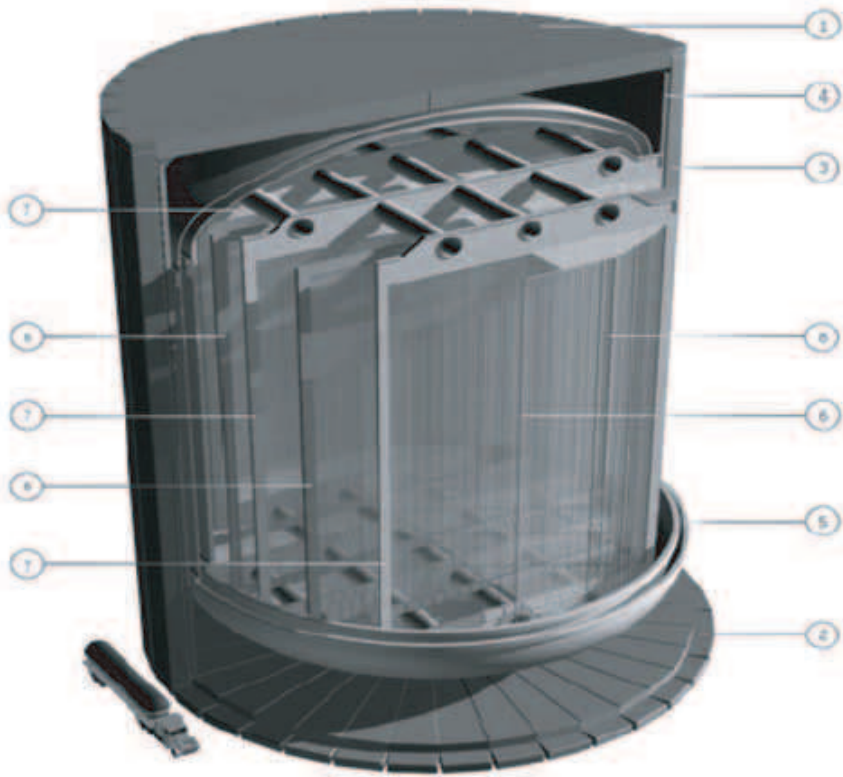,width=12cm}
\caption{Artistic view of the LANNDD design\protect\cite{Cline:2001pt}. See Ref. for
explanation of numbers.}
\label{fig:lanndd}
\end{figure}

We have recently started studying the effect of the magnetic field
on electrons (see Figure~\ref{eshower}) by means of full detector
simulation and reconstruction. 
Unlike muons or hadrons, the early showering of electrons 
makes their charge identification difficult. The track length
usable for charge discrimination is limited to a few radiation
lengths after which the showers makes the recognition of
the parent electron more difficult. In practice, charge discrimination
is possible for high fields:
\begin{eqnarray}
x=1X_0 \rightarrow B>0.5T\\
x=2X_0 \rightarrow B>0.4T\\
x=3X_0 \rightarrow B>0.3T
\end{eqnarray}
From the simulation, we found that the determination
of the charge of electrons of energy in the range between
1 and 5 GeV is feasible with good
purity, provided the field has a strength in the range of 1~Tesla.
Preliminary estimates show that
these electrons exhibit an average curvature 
sufficient to have electron charge discrimination better than
$1\%$ with an efficiency of 20\%. Further studies are on-going.

In parallel, we have initiated an R\&D to test liquid argon imaging in a magnetic
field\cite{laffranchi} (See Figure~\ref{fig:laffranchi}). We have built a small liquid argon TPC (width 300~mm, height 150~mm, drift
length 150~mm) and placed
it in the recycled SINDRUM-I magnet\footnote{The magnet was kindly lend to us by PSI, Villigen.}
which allows us to test fields up to 0.5~T. The ongoing test program includes (1) checking
the basic imaging in B-field (2) measuring traversing and stopping muons (3) test charge
discrimination (4) check Lorentz angle. Results are expected in 2004.

The design of a magnetized liquid argon TPC of 70~kton has been considered\cite{Cline:2001pt}.
It is based on a direct extrapolation (``scaling up'') of the technique developed by the ICARUS
Collaboration, however, embedded in a very large magnet (See Figure~\ref{fig:lanndd}).
A magnetized liquid argon detector would offer unequaled physics opportunities at a neutrino
factory\cite{Bueno:2001jd,Bueno:2000fg}. However, the technical feasibility of the enormous
magnet with its gigantic yoke remains an unsolved challenge. Further engineering studies
are mandatory before this technique can be proposed in an experiment of this scale.

\begin{figure}[htb]
\centering
\epsfig{file=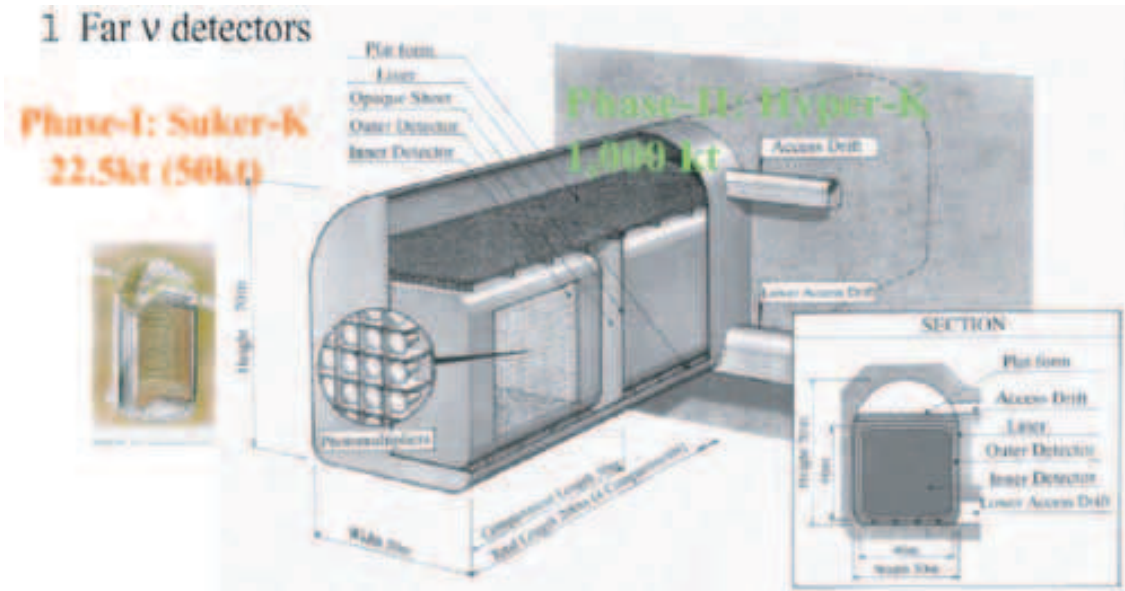,width=15cm}
\caption{Super- and Hyper-Kamiokande (see Ref.\protect\cite{Itow:2001ee}).}
\label{fig:hyperk}
\end{figure}

\section{Water  versus Liquid Argon}
Both water Cerenkov and liquid argon detectors will cover a broad
physics program, including the observation of atmospheric neutrinos, 
solar neutrinos, supernova neutrinos, and search for
proton decays, in addition to the accelerator physics program.

Giant water detectors in the range of a megaton are perceived as a ``straightforward'' extrapolation
of existing detectors like SuperKamiokande\cite{Totsuka:mc}. There have been several proposals,
as an example the design of the Hyper-Kamiokande\cite{Itow:2001ee} considered in the context
of the JHF program is reported in Figure~\ref{fig:hyperk}.

On the other hand, the liquid argon technique is considered as a difficult or unsafe, an almost
impractical, solution for large detectors. In this section, we discuss briefly the
physical properties of water and liquid argon. These are summarized in Table~\ref{tab:watlar}.
The densities of the two liquids are very similar, with a 40\% advantage for
liquid Argon. The radiation and interaction lengths are rather similar, even though
liquid Argon has roughly half the radiation length than water. The stopping power
$dE/dx$ of particles is similar in water and liquid Argon. 
Hence, neutrino
interactions or other rare events will develop very similarly in liquid argon or water!

The refractive index in the visible
spectrum is 1.33 for water and 1.24 for liquid Argon. This means that the Cerenkov
emission properties of both media are very similar. Indeed, the Cerenkov angle $\theta_C$
is $42^o$ in water and $36^o$ in liquid argon. The number of Cerenkov photons
produced per track unit length and per unit photon energy is\cite{pdg}
\begin{eqnarray}
\frac{d^2N}{dEdx} & = & \frac{\alpha}{\hbar c}\sin^2\theta_C \\
& \approx & 370 \sin^2\theta_C\ \ \rm eV^{-1} cm^{-1}
\end{eqnarray}
Accordingly, it is $\approx 160\rm\ eV^{-1} cm^{-1}$ for water and  $\approx 130\rm\ eV^{-1} cm^{-1}$
for liquid Argon. Hence, both liquids have similar Cerenkov imaging capabilities. 
Cerenkov light from penetrating muon tracks has been 
successfully detected in a liquid Argon TPC\cite{10m3cerenkov}.
 
\begin{table}[tb]
\begin{center}
\begin{tabular}{|l|c|c|}
\hline
Property & Water & Liquid Argon \\
\hline
Density ($g/cm^3$) & 1 & 1.4 \\
Radiation length (cm) & 36.1 & 14.0 \\
Interaction length (cm) & 83.6 & 83.6 \\
$dE/dx$ (MeV/cm) & 1.9 & 2.1 \\
Refractive index (visible) & 1.33 & 1.24 \\
Cerenkov angle & 42$^o$ & 36$^o$ \\
Cerenkov $d^2N/dEdx$  $\rm (eV^{-1} cm^{-1})$ & 160 & 130 \\
Muon Cerenkov threshold (MeV/c) & 120 & 140 \\
Scintillation  $\gamma/MeV$ @ $E=0$ & No & Yes, $\approx 5\times 10^3$ \\
Electron mobility & & 500 $cm^2/Vs$ \\
Long electron drift path & no & possible \\
\hline
\end{tabular}
\caption{Comparison between water and liquid Argon.}
\label{tab:watlar}
\end{center}
\end{table}

\begin{figure}[htb]
\centering
\epsfig{file=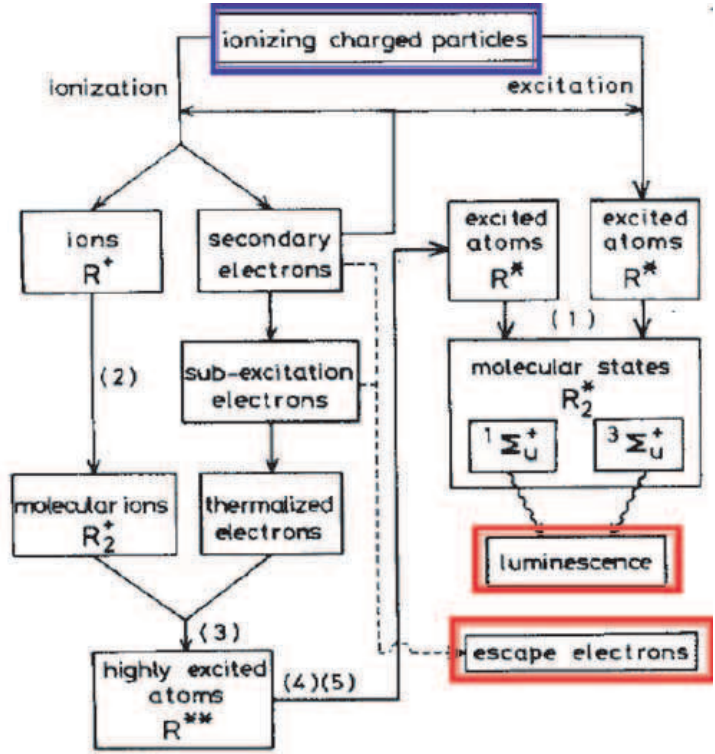,height=10cm}
\caption{Processes induced by charged particles in liquid noble gases (From Ref.\protect\cite{suzukilar}).
For particle with velocity $\beta>1/n$ there is also Cerenkov light emission (not included in the chart).}
\label{fig:suzukilar}
\end{figure}

An advantage of noble liquid gases compared to water is their very high
scintillation light (luminescence)
yield, comparable to that of $NaI$ crystals. 
The average energy needed
to produce a photon is defined as $W_\gamma$. Its measured value (at zero electric field)
is 19.5~eV for liquid Argon, 52~eV for liquid Krypton and 38~eV for liquid Xenon.
This scintillation is produced\cite{doke,suzukilar} by the de-excitation of a dimer molecule 
$(Ar-Ar^+)^*$ by the emission of a single UV photon (see Figure~\ref{fig:suzukilar}). 
In the case of liquid Argon,
scintillation appears essentially as a monochromatic line at a wavelength $\lambda = 128\rm \ nm$.
The produced light via this mechanism is not energetic enough to further ionize,
hence, the medium is transparent to its scintillation. 
In addition, it should
be stressed that in the liquid argon technique the medium is purified to very high levels
(e.g. less than 1 in $10^{10}$ oxygen-equivalent impurities) in order to ensure drifts of
electrons over long distances without impurity attachment losses,  hence it has excellent light
transmission capabilities. The propagation of light is dominated by Rayleigh scattering.
The scattering length is computed to be about 90~cm\cite{Seidel:2001vf}
 for the scintillation line at $\lambda \simeq 128~nm$.
Owing to the $\lambda^4$ dependence of the Rayleigh scattering process, we expect no effect
for visible light over distances of 100's of meters. Visible photons therefore travel the medium
in straight lines essentially unperturbed. It is also possible to increase the apparent
scattering length of the scintillation light by doping the liquid argon with xenon (typ. 5\% mixture)
which shifts the light to longer wavelengths.

\section{Combining Cerenkov, scintillation light and charge readout in a liquid Argon TPC}

Since water and liquid Argon have very similar Cerenkov light emission
properties and also similar physical properties in terms of radiation lengths,
interaction lengths, etc.. the events in water and in liquid Argon look very
much the same and the techniques developed in Kamiokande and Superkamiokande\cite{Totsuka:mc}
for the reconstruction and analysis of events can be readily ``transposed''
to the liquid argon case. Hence, the performance of a liquid argon detector
with Cerenkov light readout is at least equivalent to that of a giant water
Cerenkov detector from the point of view of event detection, reconstruction
and analysis.

\begin{figure}[htb]
\centering
\epsfig{file=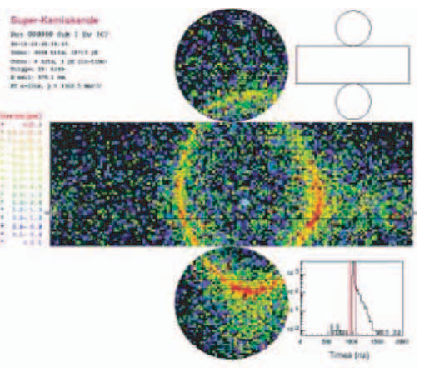,width=7cm}
\epsfig{file=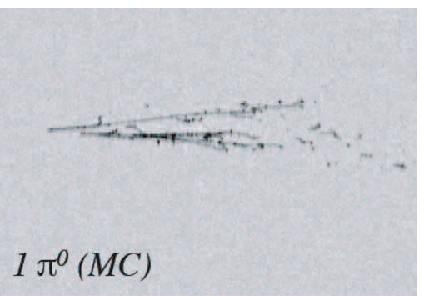,width=7cm}
\caption{Neutral pion reconstruction via (a) Cerenkov (left) (b) charge imaging TPC (right) .}
\label{fig:pizero}
\end{figure}

Of course, the overall performance of the liquid argon TPC profits
greatly from its tracking imaging properties not available in Cerenkov imaging.
The non-destructive, multiple plane readout allows
to reconstruct images in space.
With the imaging quality of a bubble-chamber 
all particles can be fully reconstructed in space, with extremely low thresholds, and the excellent
calorimetry allows to reconstruct energies very precisely. 
Tracking and calorimetry
provide momenta, particle identification, clean $e/\pi^0$ separation, etc.  Figure~\ref{fig:pizero}
shows for example the reconstruction of a neutral pion via Cerenkov and charge imaging.
Multiple showers are difficult identify in water as their Cerenkov rings tend to easily overlap.

The most striking difference between water Cerenkov and liquid argon imaging is illustrated
in Table~\ref{tab:certhres}, where the Cerenkov emission momentum threshold and the corresponding
range in liquid argon are listed
for various particles. For example, a proton becomes visible in water Cerenkov detector when
its momentum is greater than 1070~MeV. At this momentum, a proton has a range of about 80~cm
in liquid argon! With a typical wire pitch of 3~mm, particles like kaon, protons, etc. can be well
detected down to very low momenta. 

\begin{table}[tb]
\begin{center}
\begin{tabular}{|c|c|c|}
\hline
Particle&Cerenkov thr. in H$_2$O (MeV/c) &Range in LAr (cm)\\
\hline
$e$&0.6&0.07\\
$\mu$&120&12\\
$\pi$&159&16\\
$K$&568&59\\
$p$&1070&80\\
\hline
\end{tabular}
\caption{Momentum threshold for Cerenkov light emission and corresponding
range in liquid Argon for various particles.}
\label{tab:certhres}
\end{center}
\end{table}

\begin{figure}[htb]
\centering
\epsfig{file=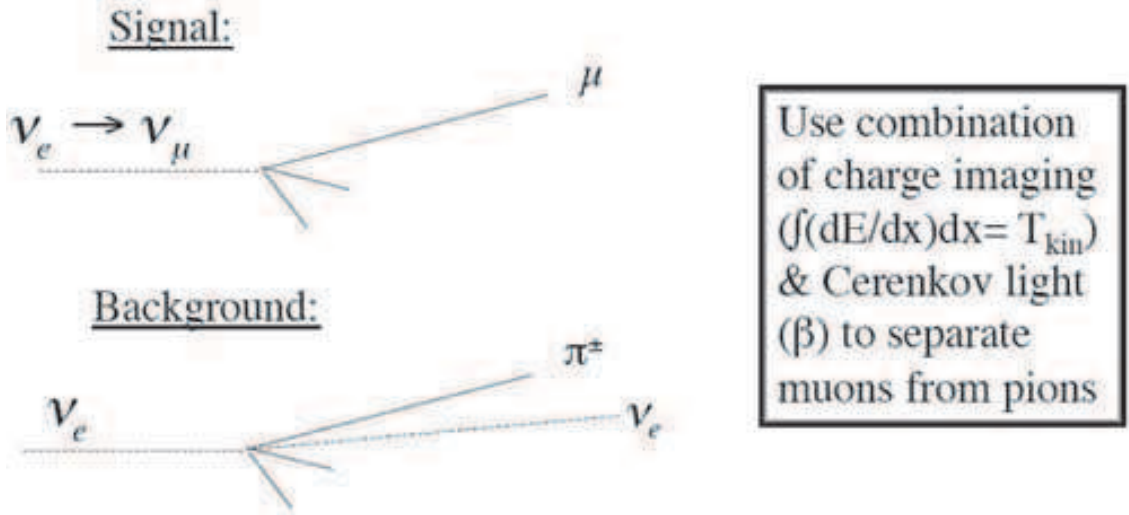,width=15cm}
\caption{Illustration of signal and background given by pion/muon confusion in 
beta-beams, when searching for $\nue\rightarrow\numu$ oscillations.}
\label{fig:mupisepar}
\end{figure}

How can one then profit from the readout
of the Cerenkov light in addition to the imaging? The combination of the information
from the tracking, energy (e.g. $dE/dx$ and kinetic energy) with the Cerenkov
light provides improved particle identification. In particular, one
can in this way separate pions from muons, a very important tool in the context
beta-beams, as illustrated in Figure~\ref{fig:mupisepar}.

\begin{figure}[htb]
\centering
\epsfig{file=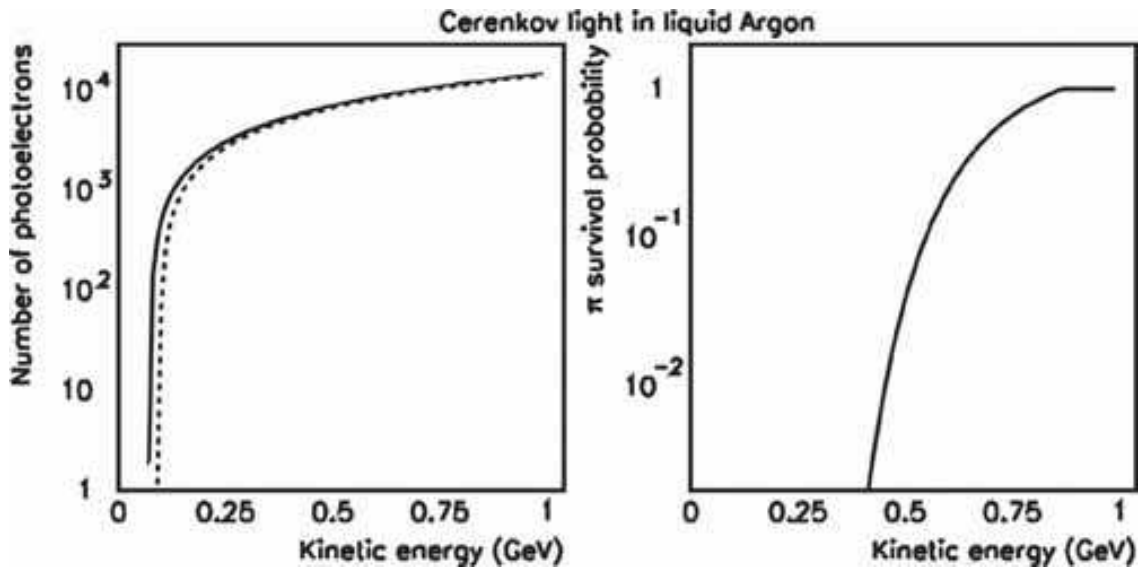,width=15cm}
\caption{(Left) Number of photoelectrons between 160 and 600~nm emitted
by muons (line) and pions (dashed). (Right) charged pion survival probability for 90\% muon acceptance efficiency.
Both quantities are plotted as a function of the kinetic energy of the particle. We assumed a 20\% coverage and
a 20\% quantum efficiency. For the pion, we assumed that no hadronic interaction takes place along
its range. In reality, the observation of an interaction via the imaging suppresses further pions versus
muons.}
\label{fig:pimusepa}
\end{figure}

The number of photoelectrons between the wavelengths of 160 and 600~nm (the typical
acceptance of a PMT) emitted
by muons and pions along their trajectory before they range out 
is shown in Figure~\ref{fig:pimusepa}(left)  as a function of the kinetic
energy of the particle. We assumed a 20\% coverage and
a 20\% quantum efficiency and that photoelectron counting is possible.
In the kinetic region of interest, the number of photoelectrons varies
from 1000 to 10000. Pions produce as expected slightly less photons.
The statistical error is so precise that it should allow separation between the two
hypotheses. For the pion, we assumed that no hadronic interaction takes place along
its range. In reality, the observation of an hadronic interaction via the imaging can be
used to discriminate pions versus muons. The charged pion survival
probability for 90\% muon acceptance efficiency is plotted in Figure~\ref{fig:pimusepa}(right).
In the range between 100~MeV (threshold) and 850~MeV, 
the method is effective at separating muons from pions.

\section {A Giant Liquid Argon TPC with charge imaging, scintillation and Cerenkov light readout}
Although the liquid argon TPC technology has been demonstrated to be mature, 
the possibility to construct a giant liquid argon TPC remains for many
an impossible technical task. In this section, we describe some issues
that to our mind show that giant liquid argon detectors might be technically
feasible.

The ICARUS collaboration has proposed an underground modular T3000 detector
for the LNGS laboratory based on the cloning of the T600 detector\cite{t3000}. The T3000 would
be composed of T600 + T1200 + T1200. The design is fully proven by the successful
technical run on the surface. Further modules are ready to be built by industry.
A 10~kton detector based on this design ``could'' be ordered today, even though this
would not be most optimal financially. Following a successful scaling up strategy, one could
envision building bigger supermodules based on the ICARUS-Airliquide
technique, by readily increasing the dimensions of the planned ICARUS T1200 by a factor 2 
in each directions: this would yield $(2^3)\times T1200 \approx T10K$. Hence, it 
seems conceivable to scale up the ICARUS dewar to a 10 kton volume. However,
to reach a total mass of 100~kton would still require a large number
of such supermodules (in fact, $10\times 10~kton \approx 100~kton$, note however that
the fiducial volume is less given the modularity). The modularity increases the complexity
of the system.

It appears from the above discussion that contrary to a modular approach, a single giant
volume is the most attractive solution.  In fact, it appears that the maximum size of the single
module is limited by the requirement
to locate the detector underground in a cavern and not by the possibility to build
a large cryogenic tanker of the needed size. Is a strong
R\&D program required to extrapolate the liquid Argon TPC to the 100~kton scale?
Or can it be achieved in say one (or two) step(s)? In the following, we try to address
the feasibility of a single volume 100~kton liquid argon detector.

\subsection{Overview of the basic design parameters}
\begin{figure}[htb]
\centering
\epsfig{file=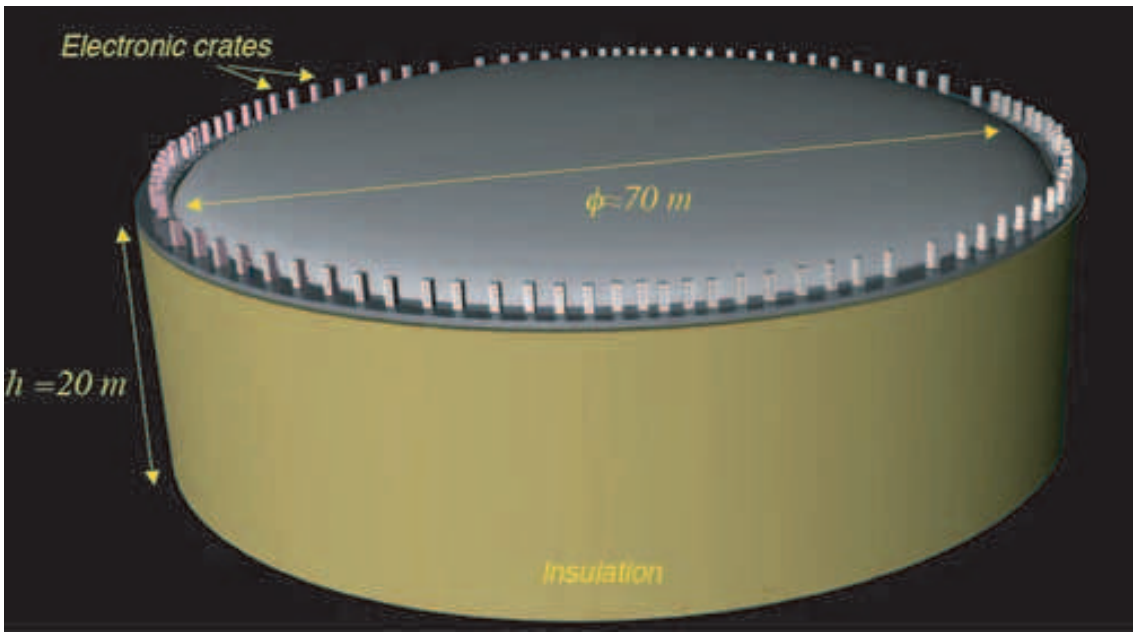,width=15cm}
\caption{An artistic view of a 100~kton single tanker liquid argon detector. It appears
that the feasibility of a volume of this size will be limited by the requirement to find
a geologically stable underground cavern of this size. The electronic crates are located
at the top of the dewar. }
\label{fig:t100k3d}
\end{figure}

An artistic view of the detector is shown in Figure~\ref{fig:t100k3d}. A summary
of parameters are listed in Table~\ref{tab:sumpar}.
The detector can be mechanically subdivided into two parts: (1) the liquid argon tanker and (2) the inner detector instrumentation.
For simplicity, we assume at this stage that the two aspects can be decoupled. 

The basic design parameters can be summarized as follows:
\begin{enumerate}
\item Single 100 kton ``boiling'' cryogenic tanker with Argon refrigeration (in particular,
the cooling is done directly with Argon, e.g. without nitrogen)
\item Charge imaging + scintillation + Cerenkov light readout for complete event information
\item Charge amplification to allow for extremely long drifts: the detector is running in bi-phase mode.
In order to allow for long drift ($\approx 20\rm\ m$), 
we consider charge attenuation along drift and compensate this effect with 
charge amplification near anodes located in gas phase.
\item Absence of magnetic field
\end{enumerate}

\begin{table}[htb]
\caption{
\bf Summary parameters of the 100 kton liquid Argon detector
} 
\small \begin{tabular}{|p{0.4\linewidth}|p{0.5\linewidth}|}
\hline
Dewar&$\Phi\approx$ 70~m, height $\approx 20~m$, passive perlite insulated, 
heat input $\approx 5$~W/m$^2$ \\
\hline
Argon Storage & Boiling argon, low pressure ($<$~100~mb overpressure)\\
\hline
Argon total volume & 73118 m$^3$ (height = 19 m),ratio 
area/volume$\approx$15\%\\
\hline
Argon total mass&{\bf 102365 TONS} \\
\hline
Hydrostatic pressure at bottom&$\approx 3$~atm\\
\hline
Inner detector dimensions & Disc $\Phi \approx$ 70~m located in gas 
   phase above liquid phase\\
\hline
Electron drift in liquid &20~m maximum drift, HV= 2MV for E=1~kV/cm,
v$_d\approx 2 $~mm/$\mu$s, max drift time $\approx$~10~ms \\
\hline
Charge readout views& 2 independent perpendicular views, 3~mm pitch, in gas
phase (electron extraction) with charge amplification\\
\hline
Charge readout channels&$\approx 100000$\\
\hline
Readout electronics & 100 racks on top of dewar (1000
channels per crate)\\
\hline
Scintillation light readout & Yes (also for triggering), 1000 immersed
8''PMT with WLS (TPB)\\
\hline
Visible light readout & Yes (Cerenkov light), 27000 immersed 8''PMTs or
20\% coverage, single photon counting capability\\
\hline
\end{tabular}
\label{tab:sumpar}
\end{table}

\subsection{The 77'000 $m^3$ liquid argon tanker}
In order to achieve such large volumes of liquid argon, we base our design on the large industrial
expertise in the storage of liquefied natural gases (LNG, $T\simeq 110K$ at 1 bar).
The LNG technology has been developed quite
dramatically in the last decades and was
driven by the petrochemical and space rocket industries. 
The technical problems associated to their design,
construction and operation have already been addressed and solved by the petro-chemical industry. 
The current state-of-the-art contemplates tankers of
$200'000\ m^3$. 
Currently there seem to be in the world about 300 giant cryogenic
tankers with volumes larger than $30'000\ m^3$.
Large ships transporting volumes up to 145'000~$m^3$ of LNG often cross the oceans.

LNG tanks are always of double-wall construction with efficient insulation between the walls. 
Large tanks are low aspect ratio (height to width) and cylindrical in design with a domed roof. 
Storage pressures in these tanks are very low.
LNG are used when volume is an issue, in particular, for storage.

\begin{figure}[htb]
\centering
\epsfig{file=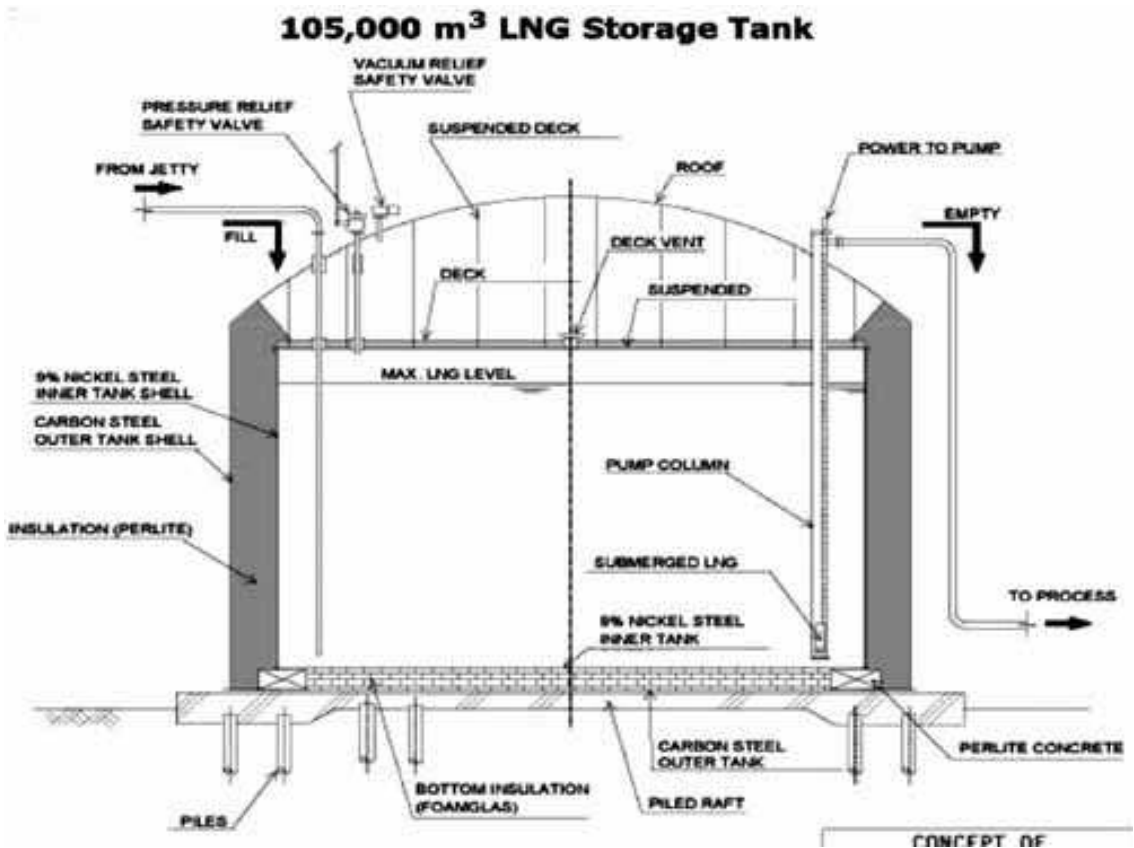,width=15cm}
\caption{Engineering design of a $105'000 m^3$ cryogenic tanker developed
by Technodyne International Limited (see Ref.\protect\cite{Technodyne}).}
\label{fig:technodyne}
\end{figure}

Of course commercial tankers are located on the surface and hence our tanker,
although of reasonable size, must face the additional constraint of being located underground.
We have contacted the Technodyne International Limited\cite{Technodyne} in the UK to initiate a feasibility
study in order to understand what are the issues related to the operation of a large
underground liquid argon detector. Technodyne is engineering company specialized
in large LNG tankers (See Figure~\ref{fig:technodyne}). 
In the baseline configuration,
we are studying the design of the standard tanker (similar to those on the surface) 
to be located underground. The tanker will be self-supporting and will not rely
on the surface of the cavern. 
Initial considerations seem to indicate that { \it the extrapolation 
from LNG to liquid Argon is rather straight-forward}. 
The geophysics of the cavern
can be understood and possible movements can be predicted for
periods of time extending to at least 30 years. The cooling of the cavern
due to heat losses is also taken into account.

\subsection{The inner detector instrumentation}

\begin{figure}[htb]
\centering
\epsfig{file=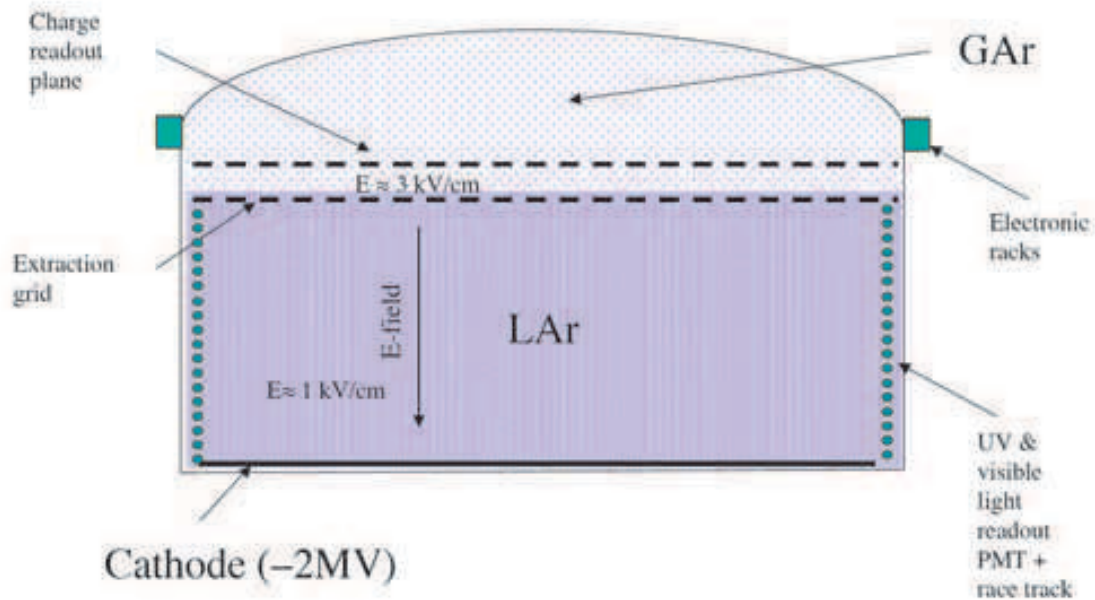,width=15cm}
\caption{Schematic layout of a 100 kton liquid Argon detector.}
\label{fig:t100schema}
\end{figure}

A schematic layout of the detector is shown in Figure~\ref{fig:t100schema}. The detector is characterized
by the extremely large volume of argon. A cathode located near the bottom of the tanker is set at $-2MV$
creating a drift electric field of 1~kV/cm over the distance of 20~m. In this field configuration ionization electrons
are moving upwards while ions are going downward.  The electric field is delimited on the sides of the tanker
by a series of ring electrodes (race-tracks) put at the appropriate voltages (voltage divider). The breakdown voltage
of liquid argon is such that a distance of about 50~cm to the grounded tanker volume is electrically safe.
For the high voltage we consider two solutions: (1) either the HV is brought inside the dewar through an appropriate
custom-made HV feed-through or (2) a voltage multiplier could be installed inside the cold volume.

The relevant parameters of the charge readout are summarized in Table~\ref{tab:readoutpar}.
The tanker contains both liquid and gas argon phases at equilibrium. Since purity is a concern for very long
drifts of the order of 20 meters, we think that the inner detector should be operated in bi-phase mode,
namely drift electrons produced in the liquid phase are extracted from the liquid into the gas phase with
the help of an appropriate electric field. Our measurements show that the threshold for 100\% efficient
extraction is about $3\ \rm kV/cm$. Hence, just below and above the liquid two grids define the appropriate
liquid extraction field. 

\begin{table}[htb]
\caption{
\bf Parameters of the charge readout
} 
\small \begin{tabular}{|p{0.4\linewidth}|p{0.5\linewidth}|}
\hline
Electron drift in liquid & 20 m maximum drift, HV=2 MV for E=1kV/cm, $v_d\simeq 2 mm/\mu s$,
max drift time $t_{max}\simeq$~10~ms\\
\hline
Charge readout views & two independent perpendicular views, 3~mm pitch\\
\hline
Maximum charge diffusion & $\sigma_D\simeq 2.8 mm$ ($\sqrt{2Dt_{max}}$ for $D=4 cm^2/s$)\\
\hline
Maximum charge attenuation & $e^{-tmax/\tau}\simeq 1/150$ for $\tau=2$~ms electron lifetime\\
\hline
Needed charge amplification & $10^{2}$ to $10^{3}$ \\
\hline
Methods for amplification & Extraction to and amplification in gas phase \\ 
\hline
Possible solutions & Thin wires+pad readout, GEM, LEM, ... \\ 
\hline
\end{tabular}
\label{tab:readoutpar}
\end{table}

\begin{figure}[htb]
\centering
\epsfig{file=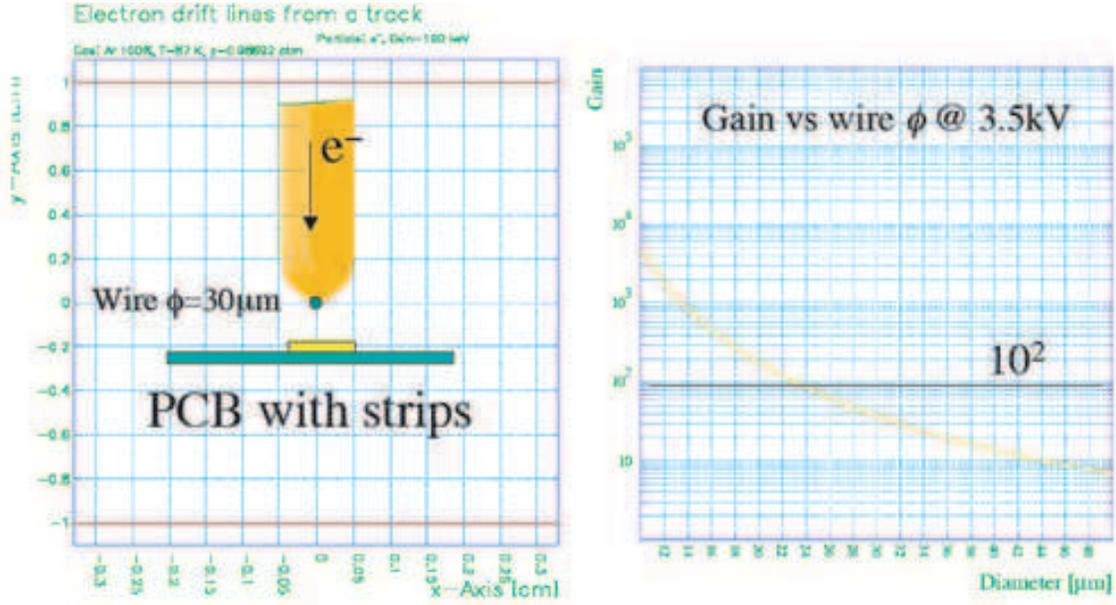,width=15cm}
\caption{GARFIELD simulation for the amplification near wires in pure argon (Ar 100\%, T=87K, p=1 atm).
(left) Geometrical setup (right) gain as a function of wire diameter. The induced signals are (1) wires and (2) strips
provide two perpendicular views.}
\label{fig:wireampli}
\end{figure}

In order to amplify the extracted charge, one can consider various options (1) amplification
near thin readout wires (like in the MPWC, see Figure~\ref{fig:wireampli}); (2) GEM\cite{Sauli:qp} or (3) LEM\cite{Jeanneret:mr}. 
Generally speaking, amplification is
technically challenging since one has to operate in pure argon and one has to face problems
of sparking, instabilities, etc. The addition of a quenching gas is not practical
since it would pollute the liquid. It would also absorb the scintillation light from the liquid argon.
Nonetheless, we have experimentally obtained encouraging results which show
that a gain of 100--1000 is achievable.
Since the readout is limited to the top of the detector, it is practical to route cables out from
the top of the dewar where electronics crates can be located around the dewar outer edges.

Amplification operates in proportional mode.
After maximum drift of 20 m at 1 KV/cm, the electron cloud
diffusion reaches approximately 3 mm which is the size of the
readout pitch. Hence, drifting along longer distances would start
degrading the quality of the images because of the diffusion
of the charge on adjacent wires. We thus think that 20 meters
corresponds to a maximum conceivable drift distance. Drifting over such distances
should be possible, allowing for some charge attenuation due to attachment
to impurities. If we assume that the reachable electron lifetime is at least $\tau\simeq 2~ms$ (this is
the value achieved on the ICARUS~T600 detector during the technical run\cite{purity} and
better values up to $10~ms$ were obtained on smaller prototypes during longer runs),
then we expect an attenuation of a factor $\simeq 150$ over the distance of 20~m. 
This loss can be compensated by the proportional gain near the anodes.

In addition to charge readout, we envision to locate PMT's around the tanker. Scintillation and Cerenkov
light can be readout essentially independently. One can profit from the ICARUS R\&D
which has shown that PMTs immersed directly in the liquid Argon is possible\cite{t600paper}. One is using commercial
Electron Tubes 8'' PMTs with a photocathode for cold operation and
a standard glass window. In order to be sensitive to DUV scintillation,
the PMT are coated with a wavelength shifter (Tetraphenyl-Butadiene).

As already mentioned, liquid Argon is a very good scintillator with about 
$50000\ \gamma/MeV$. However, this light is essentially a line at $\lambda=128\ nm$. 
Cerenkov light from penetrating muon tracks has been 
successfully detected in a liquid Argon TPC\cite{10m3cerenkov}.
This much weaker Cerenkov light (about $700\ \gamma/MeV$ between 
160~nm and 600~nm for an ultrarelativistic muon) can be separately identified with
PMT's without wavelength shifter coating, since their efficiency for the DUV light
will be very small. 

Summarizing about 1000~immersed phototubes with WLS would
be used to identify the (isotropic and bright) scintillation light. While about 27000 immersed
~8''-phototubes without WLS would provide a 20\% coverage of the surface
of the detector. As already mentioned, these latter should have single photon
counting capabilities in order to count the number of Cerenkov photons.


\subsection{Cryogenic aspects}
In order to guarantee the safety of a tanker of this size one cannot rely on
vacuum insulation. On the contrary one can use rather convential heat
insulation like for example thick layers of perlite. We envisioned an effective heat
input average over the area of the detector at the level of $5\rm\ W/m^2$. This is to be compared
to vacuum insulations which could reach $0.1\rm\ W/m^2$, however, with the risk
of a vacuum loss. 
The tanker envisaged has a very favorable ratio of area over volume of about 15\%. Hence,
the heat input of 60~kW has a small effect on the big
volume. See Table~\ref{tab:boiling}. For comparison, the refrigeration system of the
CERN LHC has a  cooling capacity  of 140~kW at 4.5~K and CERN has since 2003 
a total cryogenic capacity of 162~kW at 4.5~K.

The insulation, as efficient as it is, will not keep the temperature of LNG cold by itself. 
The liquid is stored as a ``boiling cryogen'' that is, it is a very cold liquid at its boiling point for the 
pressure it is being stored.  The liquid will stay at near constant temperature if kept at constant pressure. 
This phenomenon is called ``autorefrigeration". As long as the steam (liquid vapor boil off) is allowed to leave the 
tank, the temperature will remain constant. 
Hence, the safest way to store large quantities of cryogenic liquid is at a small
overpressure of less than approximately $200\rm\ mbar$ (basically at atmospheric
pressure) and let it evaporate. While the liquid evaporates, the temperature remains
constant. Since the process of evaporation is slow, the tanker remains in very stable and
safe conditions.

\begin{table}[htb]
\begin{center}
\caption{
{\bf Cryogenic parameters: boiling}
(Heat loss should be conservative for 3 meter thick perlite and includes
heat input from supports, instrumentation (cables), etc)} 
\small \begin{tabular}{|p{0.4\linewidth}|p{0.5\linewidth}|}
\hline
Dewar&$\Phi\approx 70$~m, height$\approx$20~m, passive 3~m thick perlite
insulated,\hfill
assumed effective heat input $\simeq$ 5 W/m$^2$\\
\hline
Total area & 12100 $m^2$\\
\hline
Total heat input & 60500 W \\
\hline
Liquid Argon evaporation rate & 0.27 liters/second or 23000 liters/day \\
\hline
Fraction of total evaporation rate & 0.03\% of total argon volume per day\\
\hline
Time to totally empty tanker by evaporation & 9 years\\
\hline 
\end{tabular}
\label{tab:boiling}
\end{center}
\end{table}

\subsubsection{Initial filling}

\begin{figure}[htb]
\centering
\epsfig{file=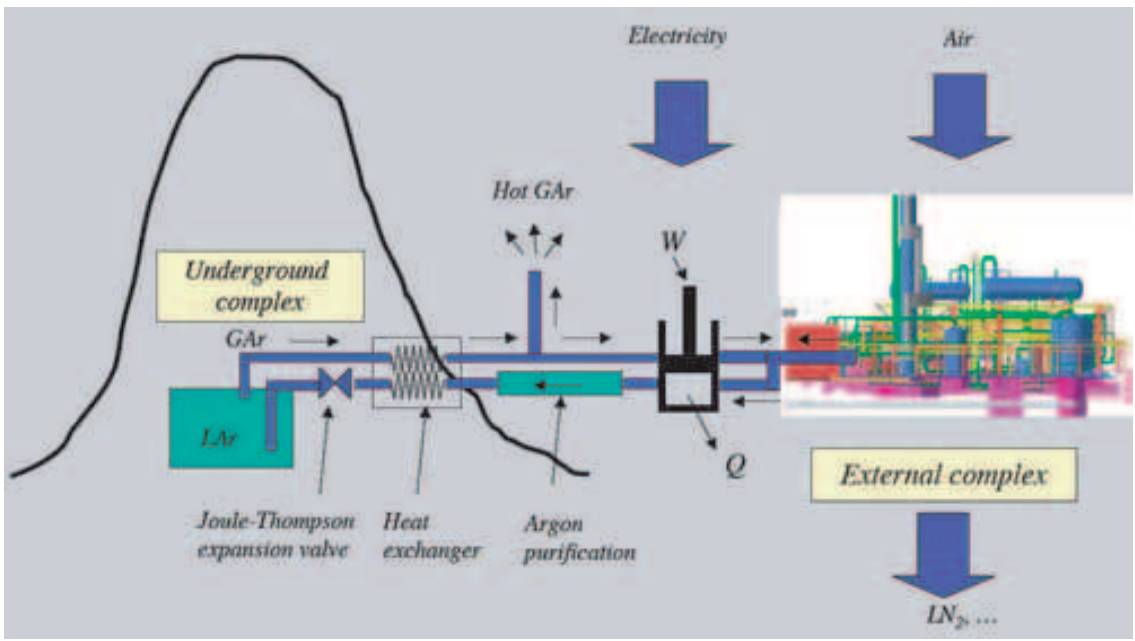,width=15cm}
\caption{Cryogenic setup of the 100 kton liquid Argon detector.}
\label{fig:t100cryo}
\end{figure}

Because of the large amount liquid argon needed to fill up the experiment (e.g. 300 ton/day to fill in 300 days),
we think that liquid argon could be produced locally. Other options (transport from various centers) are
being investigated. If we envision a dedicated cryogenic plant, then it should be located on the surface
(i.e. not underground) and connected to the detector via long vacuum-insulated pipes. See Figure~\ref{fig:t100cryo}.
Hot argon gas is compressed to high pressure (typ. 200~bar) and transported towards the underground complex.
In doing so, it can be purified. A heat exchanger is located before a standard Joule-Thompson expansion valve
which liquifies the argon into the cryogenic tanker. The gas phase is extracted from the tanker through a  heat
exchanger and transported outside. Then it can either be ventilated other fed into the compressor for a further
cooling cycle. This simple process is called a Linde-refrigerator. 

The advantage of Argon is that it is naturally available everywhere in the world, since it composes approximately 1\% of
air (by volume air is 78,1\% nitrogen, 20\% oxygen, 0.93\% argon, 0.033\% carbon dioxyde and 0.003\% rare gases
like neon, helium, krypton, hydrogen and xenon).
Liquid Argon is hence extracted from the standard process of liquefaction from air.
Air mixture is cooled down and cold gas-mixtures are separated into Oxygen, Nitrogen, Argon, ...
Liquid Argon is used to fill the experiment. The rest of the gases can be sold\footnote{However, given 
the large quantities produced
there is probably no market for it!}. An energetically efficient method is to employ the ``useless'' liquids to improve
the efficiency of the liquefiers. 

From the energy balance point of view, the Linde-refrigerator and the process of separation of Argon from air can
be ideally computed with basic thermodynamics. For a mixture of ideal gases, one has\cite{barron}
\begin{equation}
|W_s| = RT \sum_{j}y_j\ln\left(1/y_j\right)
\end{equation}
where $y_j$ is the mole fraction of the jth mixture.
It turns out that the ideal power of separation of Argon
from air is 354~kJ/kg at $T=300~K$.  Similarly, the ideal liquefaction process requires a work\cite{barron}
\begin{equation}
|W_l|=m\left(T(s_i-s_f)-(h_i-h_f)\right)
\end{equation}
where $s$ is the entropy and $h$ the enthalpy of the fluid (initial, final).
The ideal argon liquefaction is 478~kJ/kg at $T=300K$, $p=1atm$. If we assume an overall thermodynamical efficiency
of 5\%, then we find that the total electrical power needed to fill the 100~kton tanker in two years is 30~MW.
Indeed, to fill the tanker in such short time requires 150~tons of liquid argon per day! These figures are summarized
in Table~\ref{tab:filling}.

\begin{table}[htb]
\begin{center}
\caption{{\bf Cryogenic parameters: initial filling phase}
(Initial cooling of tanker not included)
} 
\label{tab:filling}
\small \begin{tabular}{|p{0.4\linewidth}|p{0.5\linewidth}|}
\hline
Liquid Argon 1st filling time  & {\bf 2 years (assumed) }\\
\hline
Liquid Argon 1st filling rate& 1,2 liters/second or 150 tons/day \\
\hline
Argon gas equivalent & 85000 m$^3$/day \\
\hline
Air volume equivalent (Ar 1\%) & 8500000 m$^3$/day $\approx$ (205
m)$^3$/day\\
\hline
Ideal power of separation of Argon mixture & 600 kW (assuming for Argon 
354~kJ/kg)\\
\hline
Ideal Argon liquefaction power  & 817kW (assuming for Argon 478~kJ/kg)\\
\hline
Assumed efficiency & 5\% \\
\hline
Estimated total plant power & {\bf 30 MW}\\
\hline
\end{tabular}
\end{center}
\end{table}

\subsubsection{Operation}

Even with a passive insulation, the favorable area to volume ratio of the considered
tanker geometry limits the evaporation of the liquid argon to 0.03\% of the volume
per day. Hence, it would take 9~years to evaporate completely the tanker by evaporation (see
Table~\ref{tab:boiling}). Once the tanker is cold and filled with argon, it will remain
therefore in very stable conditions. In order not to loose mass, one must essentially
``refill'' the amount of liquid that is lost by evaporation. This corresponds to a loss
of 0.3~liters per second.  {\it These requirements are modest compared to those
of the initial filling!}. The problem lies therefore in filling the tanker in a short time and
not to keep it cold. In order to produce new liquid argon (assuming the existing one
is simply ventilated away, a conservative assumption), one reaches with an efficiency
of 5\% a power of 6.2~MW. If the liquid Argon is circulating in closed loop and we assume
that the liquefaction power is still fully needed (conservative), then the power is 3.6~MW.
These figures are summarized in Table~\ref{tab:refill}.
\begin{table}[htb]
\begin{center}
\caption{\bf Cryogenic parameters: refilling (refrigeration)} 
\label{tab:refill}
\small \begin{tabular}{|p{0.4\linewidth}|p{0.5\linewidth}|}
\hline
Liquid Argon refilling rate &{\bf $\approx $ 0.3 liters/second } or 23000
liters/day\\
\hline
Argon gas equivalent &0.2 $m^3$/s\\
\hline
Air volume equivalent (Ar 1\%) & 20m$^3$/s or 20000 l/s \\
\hline
Ideal power of separation of argon mixture & 130kW (assuming for Argon 354 kJ/kg)\\
\hline
Ideal Argon liquefaction power  & 180kW (assuming for Argon 478 kJ/kg)\\
\hline
Assumed efficiency & 5\% \\
\hline
Estimated total power & {\bf 6.2 MW}\\
\hline
\end{tabular}
\end{center}
\end{table}

\subsection{Underground operation -- the cavern}
We believe that such a detector should be located underground in order
to provide the best possible physics output given its mass. With a shallow
depth, cosmic muons induced reactions 
which will provide dangerous backgrounds (mostly neutrons from the rock)
to rare events. We are currently studying two possible configurations: (1) a cavern
in a mine (with a vertical access through a shaft) (2) a cavern in a mountain (with
a horizontal access through a tunnel). Work is in progress and 
preliminary results from these studies should
be available soon.


\subsection{Physics program}
The physics program of a 100~kton target with the capability to study rare events
with the quality of a bubble-chamber are vast.
The physics potentialities of a giant liquid argon TPC with scintillation and Cerenkov light readout
will be reported elsewhere\cite{physglacier}. This physics program competes favorably with
a 1~Megaton water Cerenkov\cite{Itow:2001ee}. In terms of rates, one can mention:
\begin{itemize}
\item Nucleon stability: a target of $100~kton = 6\times 10^{34}$ yields a sensitivity
without backgrounds of $\tau_p/Br > 10^{34} years\times T(yr)\times \epsilon$ at the 90\%C.L.
This means that lifetimes in the range of $10^{35} years$ can be reached within 10~years
of operation. Channels like $p\rightarrow \nu K$ have been shown to be essentially background
free.
\item Atmospheric neutrinos: about 10000 atmospheric neutrinos per year and about 100 $\nu_\tau$
charged current event per year from oscillations.
\item Solar neutrinos: about 324000 events per year with electron recoil energy above 5~MeV.
\item Supernova neutrinos: a SN-II explosion at 10~kpc yields about 20000 events.
\end{itemize}
In addition, one obtains $460\ \nu_\mu$ CC per $10^{21}$ protons at 2.2~GeV (real focusing) at a distance
of 130~km, and $15000\ \nu_e$ CC per  $10^{19}$ $~^{18}Ne$ decays with $\gamma=75$ at a distance
of 130~km.

\section{Conclusions}
Given the tremendous physics potential of such detectors, we invite the community to a deep reflection 
concerning the feasibility of giant neutrino detectors and fully compare these two options:
\begin{itemize}
\item Giant 1 megaton $H_2O$ Cerenkov detector
\item Giant next-generation 100 kton liquid Argon detector, taking advantages of possible advances in the LAr TPC technology
like a bi-phase operation with charge amplification for long drift distances,
an Imaging+Scintillation+Cerenkov readout for improved physics performance, and
a Giant ÒboilingÓ cryostat (LNG technology)
\end{itemize}
These detectors offer the widest physics output (accelerator \& non-accelerator).
Coupled to the proper superbeams and beta-beams they could greatly improve our understanding of the CP-phase in the lepton sector.
International sites with proper depths and infrastructure for potentially locating such giant detectors should be reviewed and compared.
To build such large/giant detectors for only CP seems unconceivable, hence, giant detectors must have ``broad'' physics programs.
Detectors should be underground (depth to be optimized vs backgrounds).

\section{Acknowledgments}
I thank Milla Baldo Ceolin for her continuous support and
invitation to her series of  Neutrino Telescope Workshops
in the beautiful city of Venice. The help of I.~Gil-Botella and P.~Sala in the physics
simulation is greatly acknowledged. I thank P.~Picchi, F.~Pietropaolo and M.~Messina
for useful input, many discussions and great collaboration concerning charge
extraction and amplification in essentially pure argon.
The help from M.~Haworth and D.~Gurney from Technodyne International
Limited for the conception of a giant underground liquid argon tanker 
is greatly acknowledged. I thank W.~Pytel from the Polish CUPRUM
Company for providing geophysical technical information for 
an underground mine location. I thank 
A.~Zalewska for investigating mine locations in Poland. I thank L.~Mosca
for his time, help, and many fruitful discussions 
concerning the future Fr\'ejus underground laboratory. I thank A.~Bueno
for useful comments on the manuscript.


\end{document}